# Guidelines for designing 2D and 3D plasmonic stub resonators

**SOLMAZ NAGHIZADEH,[1] ŞÜKRÜ EKIN KOCABAŞ[2,*]**

[1] *Department of Physics, Koç University, Istanbul 34450, Turkey*
[2] *Department of Electrical and Electronics Engineering, Koç University, Istanbul 34450, Turkey*
*\*Corresponding author: ekocabas@ku.edu.tr*

**In this work we compare the performance of plasmonic waveguide integrated stub resonators based on 2D metal-dielectric-metal (MDM) and 3D slot-waveguide (SWG) geometries. We show that scattering matrix theory can be extended to 3D devices, and by employing scattering matrix theory we provide the guidelines for designing plasmonic 2D and 3D single-stub and double-stub resonators with a desired spectral response at the design wavelength. We provide transmission maps of 2D and 3D double-stub resonators versus stub lengths, and we specify the different regions on these maps that result in a minimum, a maximum or a plasmonically induced transparency (PIT) shape in the transmission spectrum. Radiation loss from waveguide terminations leads to a degradation of the 3D slot-waveguide based resonators. We illustrate improved waveguide terminations that boost resonator properties. We verify our results with 3D FDTD simulations.**

*OCIS codes: (250.5403) Plasmonics; (140.4780) Optical resonators; (130.7408) Wavelength filtering devices; (270.0270) Quantum Optics; (280.4788) Optical sensing and sensors*

## 1. INTRODUCTION

Surface plasmonpolaritons (SPPs) are one-dimensional bound electromagnetic waves propagating along metal-dielectric (MD) interfaces. These non-radiative modes result from longitudinal oscillation of free electrons along MD interfaces that are excited by an incident electromagnetic wave with its propagation vector modified by a prism or a grating to match the propagation vector of an SPP mode[1]. The coupling of free electrons with the incident electromagnetic wave assists the wave to propagate along metallic surfaces much further than the skin depth of the bulk metal. SPP modes exponentially decay into the dielectric and metallic media on both sides of the MD interface. This is the origin of the subwavelength confinement in the plasmonic field and more confinement can be reached by bringing two MD layers together[2]. Plasmonic devices based on SPP modes are therefore capable of overcoming the diffraction limit, which dictates the minimum size of the photonic devices. Many waveguide structures have been proposed and have been investigated for SPP waves [3-9].Among them, 2D metal-dielectric-metal (MDM) plasmonic waveguides [2] with confinement in one dimension, and 3D plasmonic slot waveguides (SWG)[7, 8] with confinement in two dimensions have gained popularity due to their high confinement, relatively long propagation length, wide bandwidth, and ease of fabrication.

Thedevelopment of resonators compatible with plasmonic waveguides helps with the design of functional devices.Stub resonators which are finite-length waveguides side coupled to the input waveguide are widely used in microwave engineering for impedance matching, filtering or switching purposes [10]. Due to the similarities between plasmonic waveguides and transmission lines, the stub idea has also been successfully extended to plasmonic applications at optical frequencies. Two-dimensional plasmonic single-stub (SS) and double-stub (DS) resonators have been employed in various applications such as tunable stop-band or band-pass filters [11, 12], low-power high-contrast switches [13], absorption switches [14], high-performance T-splitters [15], reflection-less step junctions [16], plasmonic de-multiplexers[12, 17], modulators [18], and observation of plasmonic analogue of electromagnetically induced transparency phenomenon (PIT) [19-23]. Arrays of single- or double-stub resonators have also been utilized in developing surface plasmon reflectors [24], or slow-light waveguides [14, 19].

Many of the proposed and designed plasmonic devices have used 2D plasmonic MDM waveguides as their platform for three main reasons: a) two-dimensional MDM waveguides have negligible radiation lossat bends or waveguide junctions which results in their superior performance compared to the 3D waveguides[15, 25], b) MDM waveguides are infinite in one dimension thus their 2D numerical simulations are easier to handle, c) a variety of analytical methods such as transmission line theory (TLT) [12, 16], scattering matrix theory (SMT)[11, 19], and temporal coupled mode theory (CMT) [13], have been successfully applied in the analysis of 2D MDM- based devices. Nonetheless, to fully realize an integrated plasmonic circuit and to avoid cross-talk among densely packed components, two-dimensional confinement offered by 3D SWG-based structures is needed. Additionally, SWG-based structures are much more compatible with the integrated circuit fabrication technology than theMDM-based devices. However, a slot-waveguide is inherently an open system such that the introduction of any type of discontinuity along its propagation axis results in scattering and thus radiation loss to the substrate and cladding layers [15]. Slot-waveguide propagation loss and radiation loss hinder the performance of plasmonic waveguide integrated structures. The propagation loss can be reduced by decreasing the device footprint to dimensions far less than the propagation length of the waveguide mode, however suppressing the radiation loss remains an issue to be tackled.

So far, only general properties of 2D MDM waveguides and 3D slot waveguides, such as dispersion, propagation length, and confinement factors, have been compared [8, 15]. However, to the best of our knowledge, systematic comparison between functional 2D and 3D plasmonic devices has not been made yet. Therefore, in this work we compare the performance of 2D and 3D versions of two functional plasmonic devices, i.e., single-stub resonator (SSR) and double-stub resonator (DSR), around the operating wavelength of 1550nm. In this comparison, 2D structures utilize MDM waveguides, 3D structures utilize slot waveguides.

It should be noted that some remedies have been posed to avoid radiation losses at the 3D terminated ends. For instance, in [26]stub-resonators based on three-dimensional plasmonic coaxial waveguides with performance close to their MDM counterparts have been offered.In another attempt, a slot-waveguide with a high aspect ratio cross-section has been used as the device platform [27, 28].However, fabrication of plasmonic coaxial waveguides or high-aspect-ratio structures in a manner compatible with integrated circuit technology is a challenge.

In Section 2, we introduce the 2D and 3D waveguides and single-stub and double-stub resonator geometries and the nomenclature used to define each structure.In Section 3, we overview the application of the scattering matrix model to the stub-resonators and we provide the elements of the scattering matrix required for the analysis of each stub-resonator. In Section 4, we demonstrate the simulation method for evaluating the reflection and transmission coefficients of the junctions. In Section 5, the numerical and semi-analytical results for the comparison of the 2D and 3D structures studied in this work are provided. Finally, we conclude in Section 6.

## 2. STRUCTURES

Schematics of the 2D metal-dielectric-metal waveguide and 3D slot-waveguide are depicted in Fig. 1(a) and 1(b). In both cases dielectric parts are silica with a refractive index of 1.44 and metallic parts are gold with a complex dielectric function taken from [29]. Our choice of gold is due to the fact that it is commonly used in experiments.We use a symmetric slot-waveguide with identical substrate and superstrate materials (silica) which ensures the existence of a bound mode for a broadband excitation [8]. The MDM waveguide is infinite in the x-direction and is confined only in y-direction while the slot waveguide is confined in x- and y-directions. The modes propagate along the z-direction in both geometries.

Bearing in mind the fabrication challenges of narrower gaps, we set the widths of MDM and SWG waveguides to 200nm and 220nm, respectively, and then by sweeping over the height of the SWG we found the height $h$=115nm for which the real part of the effective refractive index of the modes in the two structures were the same atthe operating wavelength of 1550nm. The real part of the complex effective refractive index for the two waveguides is shown inFig. 2. The intersection point amounts to $n_{eff}$=1.6186 at 1550nm. The propagation length, defined as $L_p=1/imag(k)=1/\alpha$ where $k=\beta+i\alpha$ is the complex propagation constant of the waveguide; for our reference MDM waveguide and SWGthis amounts to 22.5 μm and 17.5 μm at 1550nm, respectively. In all the subsequent 2D and 3D simulations the material properties and the dimensions of the waveguides will be kept fixed.There are other width and height configurations that result in the same $n_{eff}$at 1550 nm, our particular choice is based on having structures that are easy to fabricate. In microwave frequencies, the $k$value is critical for stub resonances that is why we keep$k$the same at 1550 nm for MDM and SWG waveguides.

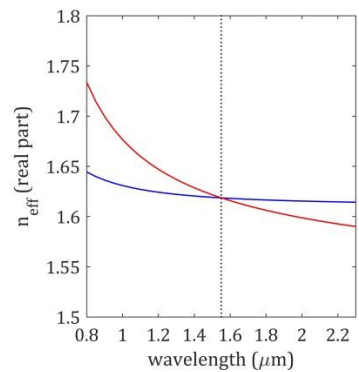

Fig.2. Effective refractive index of MDM waveguide and SWG intersecting at 1550nm (blue: MDM waveguide, red: SWG)

Schematics of the single-stub resonators (SSR) and double-stub resonators (DSR)are shown in Fig.3.In Figs. 3(b) and 3(d) the top silica cladding has been removed for a more clear visualization of the geometry. Double-stub resonators (DSR) are formed by two single-stub resonators (SSR) located at the same site along the input waveguide, indicating that SS resonators are the building blocks of the DS resonators.

## 3. ANALYTICAL MODEL

The subwavelength size of our reference waveguides compared to the operating wavelength of 1550nm ensures their single-mode operation which is the prerequisite for employing single-mode scattering matrix theory as a semi-analytical model. We applied scattering matrix theory to analyze the resonator structures.

The SS resonator is composed of three simple geometries, namely: a T-junction with input from the left, a terminated waveguide, and a T-junction with input from the top, as sketched in Fig. 4(a)-4(c). Complex reflection and transmission coefficients of these geometries are denoted as follows: $r_1$ is the reflection from the input port of the T-junction (with input from the left), $t_1$ is the transmission to its straight-output waveguide, and $t_2$ is the transmission to its cross output waveguide[Fig. 4(a)]. $r_2$ is the reflection from the terminated waveguide[Fig. 4(b)], $r_3$ is the reflection from the input port of the T-junction (with input from the top), and $t_3$gives the corresponding transmission coefficient to its cross output waveguides [Fig. 4(c)].

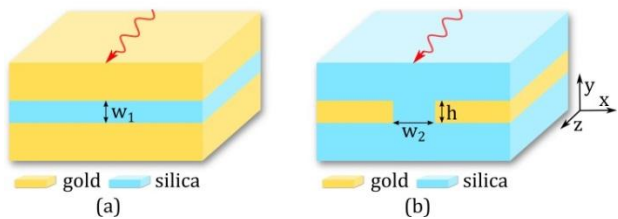

Fig.1.(a)2D MDM waveguide (gold/silica/gold) with width $w_1$=200nm, (b) SWG with dimensions $w_2 \times h$=220nm×115nm. The mode propagates in the z-direction in both cases.

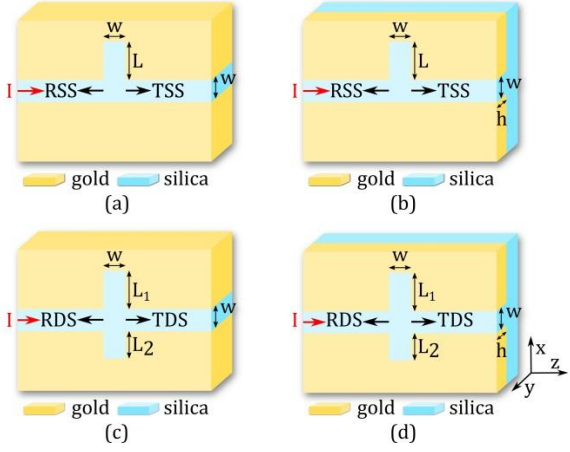
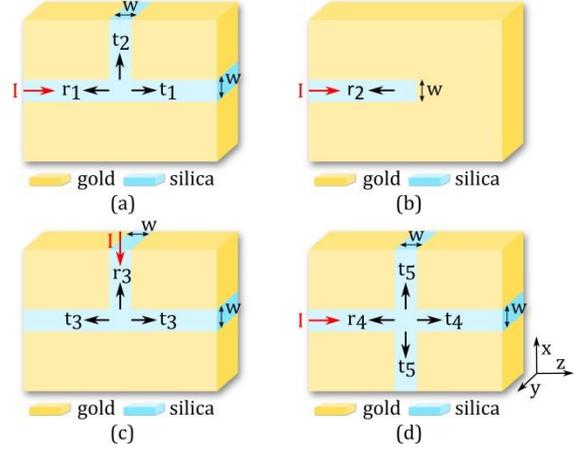

Fig.3. (a) 2D single-stub resonator, (b) 3D single-stub resonator without top cladding, (c) 2D double-stub resonator, (d) 3D double-stub resonator without top cladding.

Fig.4. (a) 2D T-junction with input from the left, (b) 2D terminated waveguide, (c) 2D T-junction with input from the top, (d) 2D X-junction.

With these definitions the complex transmission (TSS) and reflection (RSS) amplitudes of single-stub resonator with stub length $L$ are given by[11]

$$TSS = t_1 + \frac{r_2\, t_2 t_3 \exp(2ikL)}{1 - r_2\, r_3 \exp(2ikL)} \quad (1)$$

$$RSS = r_1 + \frac{r_2\, t_2 t_3 \exp(2ikL)}{1 - r_2\, r_3 \exp(2ikL)} \quad (2)$$

The first term in Eq.(1), i.e. $t_1$, is the portion of the input SPP wave that directly propagates into the straight-output waveguide of the corresponding T-junction without entering the stub [Fig. 5(a)]. However, the second term gives the contribution from the input SPP wave that enters the stub and then exits to the input waveguide. This term has been derived by the summation of all the transient portions of the incident SSP wave infinitely bouncing back and forth inside the stub to obtain the stationary response [Fig. 5(b)]. The same argument holds for Eq.(2) in which the first term $r_1$ is the reflection from the left port of the corresponding T-junction, and the second term is the portion of the SPP wave in the stub that is emitted to the left of the input waveguide.

Similarly, scattering matrix theory has been employed in the derivation of the complex transmission (TDS) and reflection (RDS) amplitudes of the 2D double-stub resonator by deconstructing the DS resonator as a combination of two geometries: a terminated waveguide and an X-junction as sketched in Fig.4(b) and 4(d), respectively. For a double-stub resonator of stub lengths $L_1$ and $L_2$ the complex transmission coefficient (TDS) and the complex reflection coefficient (RDS) are given by[14, 19]:

$$TDS = t_4 - C \quad (3)$$
$$RDS = r_4 - C \quad (4)$$

where $C = t_5^2 (2t_4 - 2r_4 + s_1 + s_2)/[t_4^2 - (r_4 - s_1)(r_4 - s_2)]$,
$s_1 = 1/[r_2 \exp(2ikL_1)]$ and $s_2 = 1/[r_2 \exp(2ikL_2)]$. Complex reflection and transmission coefficients of the X-junction are denoted as follows: $r_4$ is the reflection from the input port of the X-junction, $t_4$ is the transmission to its straight-output waveguide, and $t_5$ is the transmission to its cross output waveguide [Fig. 4(d)].

Similar to the single-stub resonator, the Eqs. (3) and (4) have been derived by assuming the incident SPP wave undergoing five different pathways depicted in Figs. 5(c)-5(g) and described as follows: Fig(c) shows the portion of the incident SPP wave that directly propagates through the X-junction, Fig. 5(d) shows the portion of the incident SPP wave that enters the upper stub, bounces there multiple times and emits out to the right of the input waveguide, Fig. 5(e) shows the portion of the incident SPP wave that enters the upper stub and bounces back and forth in the combined resonator of length $L_1+w+L_2$ and then emits out to the right of the input waveguide, similarly, Fig. 5(f) shows the portion of the input SPP wave that enters the lower stub, bounces there multiple times and emits out to the right of the input waveguide, finally Fig. 5(g) shows the portion of the input SPP wave that enters the lower stub and bounces back and forth in the combined resonator of length $L_1+w+L_2$ and then emits out to the right of the input waveguide.

## 4. SIMULATION MODEL

Equations (1)-(4) enable us to predict and investigate the properties of plasmonic single-stub and double-stub resonators without running numerical simulations. This is particularly of importance in 3D structures which require more simulation time and memory. Furthermore, the equations provide additional insight into the operational principles of SS and DS resonators.

We used the commercial software package Lumerical FDTD Solutions [30] for numerical simulations of our structures. To utilize Eqs. (1)-(4) first we need to extract all the complex reflection and transmission coefficients of the relevant 2D and 3D geometries [Fig. 4(a)-4(d)]. The methodology for obtaining complex reflection and transmission coefficients of a 3-port junction (i.e., $r_1$, $t_1$, and $t_2$) is illustrated in Fig. 6

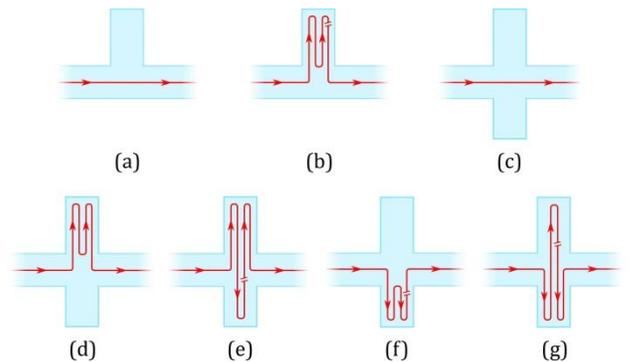

Fig.5. SPP wave scattering pathways in (a,b) single-stub resonator, (c-g) double-stub resonator.

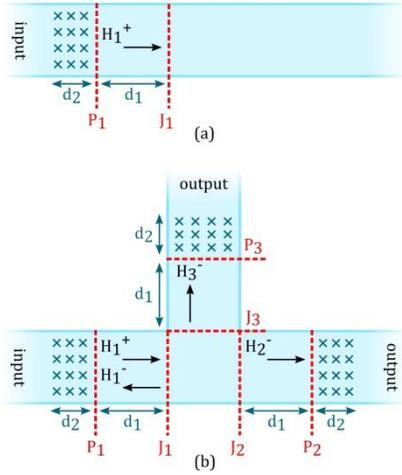

Fig.6. Evaluating (a) incident SPP wave in a straight waveguide, (b) $r_1$, $t_1$, and $t_2$, complex coefficients in a 3-port junction.

[31]. The reflection and transmission coefficients of all other geometries are extracted in a similar way.

To allow waveguide modes to completely form and to avoid higher order modes that might be excited upon reflection or scattering from junctions, in both 2D and 3D simulations SPPs are collected sufficiently away from the waveguide junctions. Therefore in both 2D and 3D simulations we set our ports ($P_1$, $P_2$, $P_3$), $d_1$=1μm away from junctions ($J_1$, $J_2$, $J_3$) and by defining a set of point time monitors (PTM) within $d_2$=0.4μm from ports (shown by the cross signs in Fig. 6) we collect time varying magnetic fields from each time monitor and then perform FFT to obtain the frequency-domain response of each PTM. We calculate the complex transmission and reflection coefficient for each point time monitor as

$$t = \frac{H_1^-}{H_1^+} \quad (5)$$

$$r = \frac{H_1^+ + H_1^-}{H_1^+} - 1 = \frac{H_1^-}{H_1^+} \quad (6)$$

Frequency-domain values of the fields are substituted into these equations and the incident SPP wave, $H_1^+$, which is used as normalization is obtained by running a separate simulation for each of the 2D and 3D straight waveguides [Fig. 6(a)]. We tested our choice of $d_1$=1 μm by increasing $d_1$ to 2 μm and observing that the $r$ and $t$ coefficients do not change their values.

In 2D simulations point time monitors are distributed in a $xy$ plane however in 3D simulations they are distributed in a $xyz$ volume. These monitors are located precisely in input and output ports such that each time monitor a distance $d$ away from the input port has an equivalent time monitor a distance $d$ away from the output port. We evaluate the reflection and transmission coefficients at all these points. We then transform the obtained coefficients to the junction locations ($J_1$, $J_2$, $J_3$). We observe that the coefficient values obtained from a collection of points at different locations in the FDTD grid [e.g. to the right of $P_2$ in Fig. 6(b)] all have very similar values in magnitude and in phase—which ensures that the simulation is setup correctly. We further average over the coefficients obtained from different grid locations to get the final values of $r$ and $t$ coefficients.

In 2D simulations we used a uniform mesh of 2nm in all directions and in 3D simulations we used a uniform override mesh of size 5nm in a $xyz$-volume surrounding input and stub waveguides and non-uniform mesh elsewhere. The mesh accuracy was set to its default value of two in 3D simulations. Since 2D simulations run quickly we set the mesh size in 2D simulations to 2nm. Nonetheless, for our 2D simulations the mesh size of 5 nm resulted in similar results as those run with 2nm mesh-size. To further check the subtleties of the PTM method we verified its results with the results obtained with FDTD Solutions built-in Mode Expansion monitors (not shown).

## 5. DISCUSSION AND RESULTS

### A. Single-stub resonators

By plugging the complex reflection and transmission coefficients of the geometries sketched in Figs. 4(a)-4(c) at 1550nm into Eq. (1), we plot the 2D and 3D $|TSS|^2$ versus stub length $L$ as shown in Fig. 7. These graphs give us the resonant orders at 1550nm and assist us in selecting the stub lengths that result in either dips or peaks in the transmission of 2D or 3D SS resonator at the operating wavelength of 1550nm. The variations in the single-stub spectrum as a function of $L$ are due to the interference of the two SPP waves undergoing different pathways: the SPP wave that goes directly through the junction and the ones that bounce within the stub [Figs. 5(a) and 5(b)]. Similarly, by plugging the relevant 2D and 3D complex reflection and transmission coefficients obtained for the wavelength interval [0.8, 2.3]μm into Eq. (1), we plot the wavelength response for the first four resonant stub lengths corresponding to a dip or a peak in the transmission spectrum at 1550nm for 2D (Fig.8) and 3D (Fig. 9) SS resonators. The specifications of the resulting SMT-predicted spectra for 2D and 3D single-stub resonators are listed in Table 1 and Table 2. To calculate the Q-factor defined as $f_0$/FWHM and the finesse defined as FSR/FWHM, where $f_0$ is the resonant frequency, FWHM is the full width half maximum of the spectrum and FSR is the free-spectral range, we plotted spectra in frequency domain to evaluate the FWHM and FSR parameters. The Q-factors of the spectra with dips have been calculated by inverting the spectra via the operation max(spectrum)-spectrum. From the information provided in these tables we see that the spectra which feature dips have less FWHM compared to the spectra that feature peaks.

We found that in both of the 2D and 3D spectra with either dips or peaks, longer stub lengths resulted in narrower FWHM but slightly lower contrast. This means that there is a tradeoff between the device footprint and its optimum performance. This observation originates from the fact that a single-stub resonator can be assumed as a Fabry-Perot (FP) resonator with partially reflecting mirrors of fixed reflection coefficients $r_1$ and $r_2$ for a specific wavelength. In a Fabry-Perot resonator the resonator FWHM linewidth is given by $\Delta \nu = c\, \alpha_r/2\pi$, where $\alpha_r$ is the effective total distributed-loss coefficient given by $\alpha_r = 1/L_p + (1/2L)\ln(1/r_1 r_2)$ [32, 33]. As a result, by increasing $L$ the FWHM decreases.

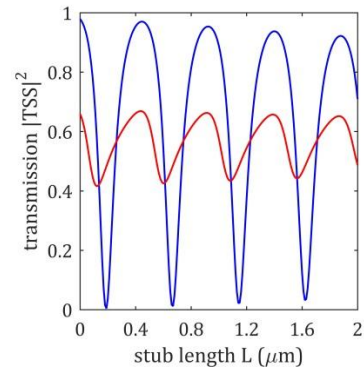

Fig.7. Power transmission $|TSS|^2$ for 2D (blue curve) and 3D (red curve) single-stub resonators versus stub length.

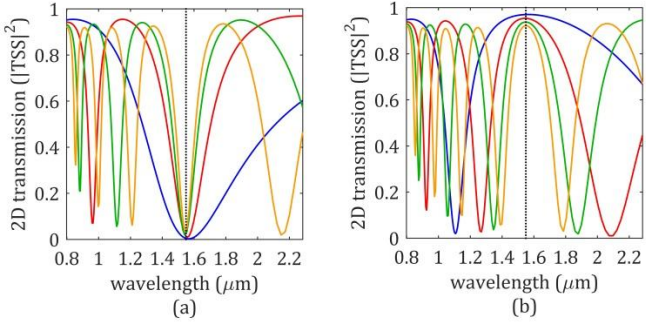
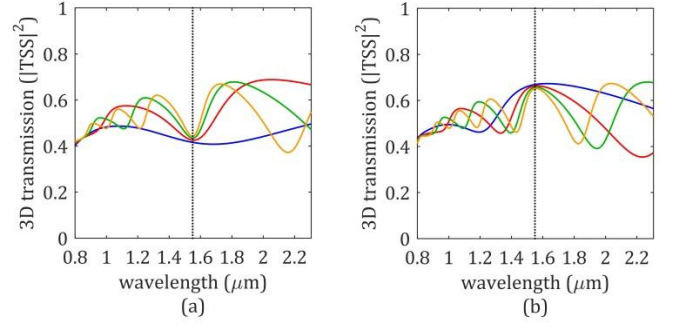

Fig.8. (a) First four resonant orders corresponding to a dip in 2D SS resonator spectrum at 1550nm (blue: *L*=190nm, red: 670nm, green: 1140nm, yellow: 1620nm), (b) first four resonant orders corresponding to a peak in 2D SS resonator spectrum at 1550nm (blue: *L*=450nm, red: 920nm, green: 1400nm, yellow: 1880nm).

Fig.9. (a) First four resonant orders corresponding to a dip in 3D SS resonator spectrum (blue: *L*=120nm, red: 600nm, green: 1080nm, yellow: 1560nm), (b) first four resonant orders corresponding to a peak in 3D SS resonator spectrum (blue: *L*=440nm, red: 910nm, green: 1390nm, yellow: 1870nm).

It is also evident (Fig.7) that the transmission contrast, i.e. the difference between minimum and maximum power levels which is a requisite for having high performance switches and filters is far less in 3D SS resonator than the 2D SS resonator. This quantity varies between 0.43 and 0.67 for 3D SS resonator while for a 2D SS resonator the correspondent value varies between nearly 0 and 0.92. Furthermore, the 2D and 3D SS resonators which are designed to be equivalent at 1550nm show peaks in their spectra for almost the same stub lengths but the dip in their spectra occurs for different stub lengths. The peak values corresponding to different stub lengths amount almost to the transmitted powers of the straight waveguides (*L*=0) in both 2D and 3D cases. However, the dip values in the 2D case almost reach zero and in the 3D case it reaches 0.43.

**Table 1. 2D Single-Stub Resonators**

| $L_{dip}$ (nm) | FWHM (nm) | Q-factor | Finesse |
|---|---|---|---|
| 190 | 800 | 2.13 | ----- |
| 670 | 270 | 5.80 | 3.60 |
| 1140 | 170 | 9.29 | 3.57 |
| 1620 | 120 | 12.52 | 3.53 |
| $L_{peak}$ (nm) | FWHM (nm) | Q-factor | Finesse |
| 450 | 1101 | 1.56 | 1.40 |
| 920 | 577 | 2.84 | 1.36 |
| 1400 | 376 | 4.23 | 1.38 |
| 1880 | 279 | 5.66 | 1.36 |

*Spectrum specifications of first four resonant orders resulting in a dip or a peak at 1550nm in 2D SS resonator power transmission.

**Table 2. 3D Single-Stub Resonators**

| $L_{dip}$ (nm) | FWHM (nm) | Q-factor | Finesse |
|---|---|---|---|
| 120 | ----- | ----- | ----- |
| 600 | 467 | 2.99 | ----- |
| 1080 | 260 | 5.69 | 2.25 |
| 1560 | 182 | 8.24 | 2.32 |
| $L_{peak}$ (nm) | FWHM (nm) | Q-factor | Finesse |
| 440 | ----- | ----- | ----- |
| 910 | 907 | 2.10 | 0.95 |
| 1390 | 550 | 3.18 | 1 |
| 1870 | 400 | 4.22 | 1 |

*Spectrum specifications of first four resonant orders resulting in a dip or a peak at 1550nm in 3D SS resonator power transmission.

The low contrast and the shift in dip resonances in the 3D case can be explained by understanding the operational principle of the single-stub resonator. To understand the operational mechanism of single-stub resonators, in Fig. 10(a) we show the transmitted power of a 2D single-stub resonator versus stub length *L* at 1550nm (blue line) along with two different phase plots, all obtained by scattering matrix theory: a) phase difference between the SPP wave passing directly through the junction and the SPP wave coming back from the stub (yellow line) which in fact is the phase difference between the first and second terms of the Eq. (1), b) round-trip phase shift (red line) that the SPP wave entering the stub gains in the stub given by: $\varphi_{r_2}(\lambda) + 2k(\lambda)L + \varphi_{r_3}(\lambda)$, where $\varphi_{r_2}(\lambda)$ is the phase shift that the SPP wave experiences upon reflection from the terminated end of the stub, $\varphi_{r_3}(\lambda)$ is the phase shift experienced by the SPP wave as it reflects from the T-junction with input from the top, and $2k(\lambda)L$ is the round-trip phase shift that the SPP wave gains as it propagates along a stub of length *L*.

The phase analysis of the two SPP waves propagating along two different pathways in a single-stub resonator [Figs.10(a) and 10(b)] reveals that when the phase difference between the two SPP waves (yellow line) is $2\pi(\pi)$ there is a peak (dip) in the transmitted power. Moreover, when there is a dip in the transmission the round-trip phase gained by the SPP wave inside the stub is $2\pi$. Therefore, in the formation of a dip, the SPP wave that enters the stub undergoes constructive interference and bounces multiple times inside the stub before coming out. In this case due to the open nature of the 3D SSR the SPP wave suffers more from the radiation loss compared to the case when it does not undergo constructive interference inside the stub. Consequently, the attenuated SPP wave leaving the stub has smaller amplitude compared to the SPP wave that directly passes through the argument holds for the case when we have a peak in the transmission. The peak transmission in the 3D case is lower compared to the 2D case because the SPP wave that enters the stub loses almost half of its power in the first round, which causes poor constructive interference between the two SPP waves that undergo different pathways despite being in phase.

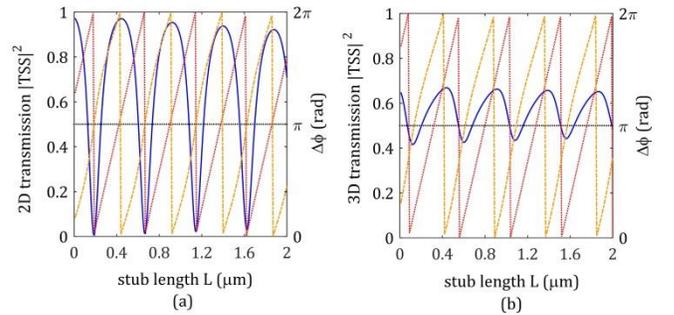

Fig.10. Relation between the variation of the transmission amplitude (blue line) and the evolution of phases (red and yellow lines) versus stub length *L* in (a) 2D SS resonator, (b) 3D SS resonator.

The main reason behind the shift in the resonant lengths corresponding to dips in 3D case with respect to the 2D case is due to the fact that, when there is a dip, the SPP wave circulating in the stub satisfies the resonance condition given by $\varphi_{r_2}(\lambda) + 2k(\lambda)L + \varphi_{r_3}(\lambda) = 2m\pi$ ($m = 0, 1, 2, \dots$) [red line in Fig. 10(a)]. Comparison of the phases of $r_2$ and $r_3$ coefficients in 2D and 3D cases (not shown) reveals that in 3D, $\varphi_{r_2}(\lambda) + \varphi_{r_3}(\lambda)$ contribution is larger compared to 2D, hence a shorter $L$ value suffices to reach the resonance.

Another point is that the resonant dips and peaks that occur at wavelengths other than 1550 nm in the SMT-predicted spectra (Fig. 8 and Fig. 9) do not overlap as the resonance order increases. By changing the stub length the free-spectral-range (FSR) and accordingly the location of the resonances change. In ideal FP-resonators that are all resonant at a given wavelength $\lambda_0$, the resonators with higher resonance orders are expected to fit an integer amount of peaks/dips in between the peaks/dips of the resonator with the lowest order. However, in our case the resonance condition has the general form $\varphi_{r_2}(\lambda) + 2k(\lambda)L + \varphi_{r_3}(\lambda) = m\pi$ ($m = 0, 1, 2, \dots$) with dispersive $k$ and $\varphi_{r_2}(\lambda)$ and $\varphi_{r_3}(\lambda)$ values which leads to deviations from the ideal FP model.

We verified the prediction of scattering matrix theory with numerical FDTD simulations. We simulated 2D and 3D SS resonators with stub lengths both set to the third resonant length of the 3D SS resonator at 1550nm, i.e. $L$=1.39μm as predicted by SMT in Table 2. The simulation results shown in Figs. 11(a) and 11(b), verify the predictions of SMT and its applicability to the 3D structures.

The obtained semi-analytic and numeric results reveal the low capability of 3D SS resonator for being used in switching and filtering purposes compared to 2D SS resonators. This is mostly due to the low contrast of the 3D SS resonator transmission spectrum. A couple of solutions have been offered to either reduce the radiation losses at the 3D terminated ends and thus increase the contrast of a 3D SS resonator or design a slit-like 3D plasmonic filter. For instance, in [34] stub region has been filled with a high refractive index material. This increases the confinement of the mode in the stub region and the contrast of the transmission spectrum. Another way of increasing the contrast of the 3D SS resonator is to increase the height of the slot-waveguide metallic layer to 1-3μm while keeping its width as small as 50 nm [27, 28]. The narrower gap for 3D slot waveguides provides more confinement and improves device performance. To get around the radiation losses, a 3D slit-like plasmonic filter in thin metallic layers has been designed and has been shown to be more successful than the 3D SS resonator [28, 35].

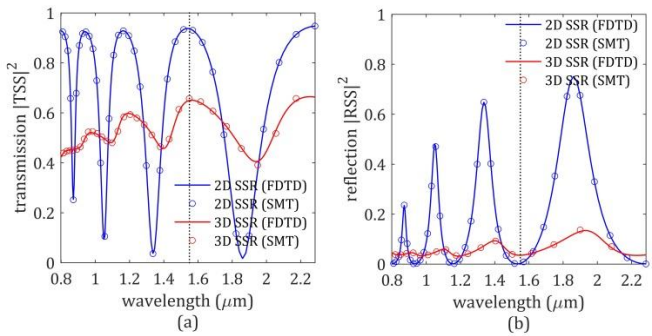

Fig.11. (a) Transmission; (b) reflection coefficients of 2D and 3D SS resonators with stub length $L$=1390nm predicted by scattering matrix theory and verified with numerical FDTD simulations.

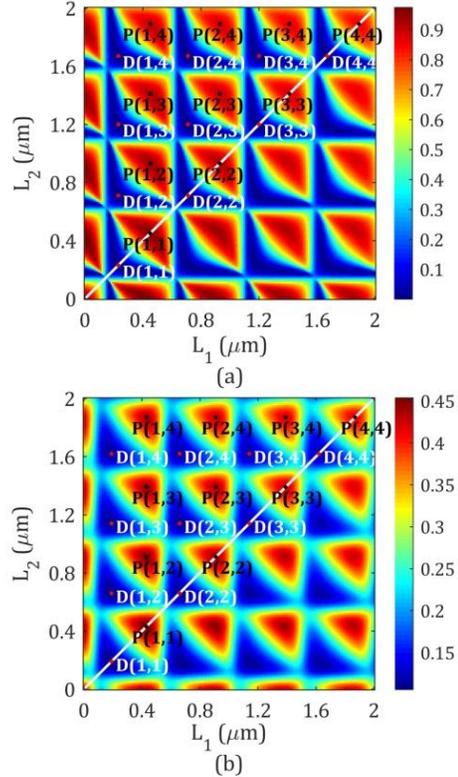

Fig.12. Power transmission $|TDS|^2$ vs $L_1$ and $L_2$ at 1550nm for the (a) 2D DS;(b) 3D DS resonator.

**B. Double-Stub Resonators**

Similar to the single-stub resonator, by substituting the complex reflection and transmission coefficients at 1550nm of the relevant geometries [Figs.4(b) and 4(d)] into Eq.(3) we plot the power transmission coefficient, $|TDS|^2$, for the 2D and 3D double-stub resonators versus stub lengths $L_1$ and $L_2$ both ranging from 0 to 2μm as shown in Figs. 12(a) and 12(b), respectively.

These transmission maps depict the resonant orders for which there is either a dip or a peak in the power transmission spectrum of double-stub resonators and thus are helpful in choosing the appropriate stub lengths depending on the application.

Due to the symmetry of the geometry, we focus only on the length pairs that lie above the $L_1$=$L_2$ line, shown by the white line. The resonant orders corresponding to peaks (dips) in transmission are denoted by black (red) dots and the letter P (D).

In the double-stub resonators the incident SPP wave has the opportunity to propagate through five different pathways as illustrated in Figs. 5(c)-5(g). Variation of the power transmission versus stub lengths originates from the fact that for some length pairs the five SPP waves interfere constructively (destructively) to form a peak (dip) in the resulting power transmission spectrum.

Moreover, in the $|TDS|^2$ plots, $L_1$=0 and $L_2$=0 axes, which due to symmetry are identical, give the power transmission spectrum of the single-stub resonator at 1550 nm with stub length ranging from 0 to 2μm. It is clear that along these axes, the length interval for which we have a dip in the transmission spectrum of the SS resonator is less than the length interval for which we have a peak. As tuning of stub length is junction which results in a poor destructive interference between the two wave sdespite being out of phase and thus low contrast. A similar analogous to the tuning of wavelength this further proves that spectralFWHM of the SS resonator, which features a dip in transmission, is narrower than the one that features a peak in the transmission.

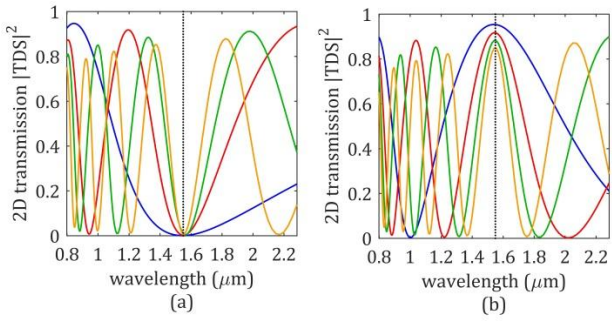

Fig.13. (a) Blue line: $L_1=L_2$=0.23µm, red line: $L_1=L_2$=0.71µm, green line: $L_1=L_2$=1.2 µm, yellow line: $L_1=L_2$=1.67 µm. (b) blue line: $L_1=L_2$=0.45µm, red line: $L_1=L_2$=0.93µm, green line: $L_1=L_2$=1.41µm, yellow line: $L_1=L_2$=1.89µm. Black dotted line specifies the location of the operating wavelength 1550nm.

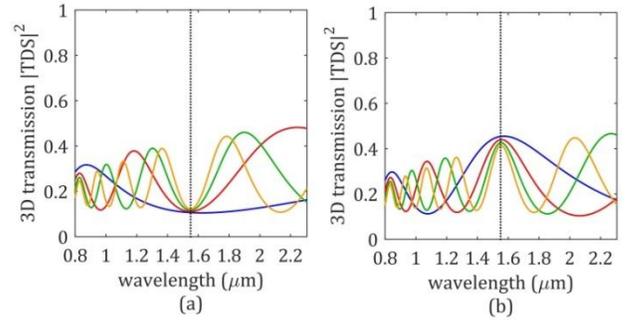

Fig.14. (a) Blue line: $L_1=L_2$=0.19µm, red line: $L_1=L_2$=0.66µm, green line: $L_1=L_2$=1.14µm, yellow line: $L_1=L_2$=1.62 µm. (b) blue line: $L_1=L_2$=0.43 µm, red line: $L_1=L_2$=0.91um, green line: $L_1=L_2$=1.39µm, yellow line: $L_1=L_2$=1.879µm. Black dotted line specifies the location of the operating wavelength 1550nm.

In Figs. 13(a) and 13(b) we plot the SMT-predicted spectra of the 2D double-stub resonator $|TDS|^2$for the first four resonant orders of equal stub lengths that result in a peak [denoted as P(1,1), P(2,2), P(3,3), P(4,4) in Fig. 12(a)] or a dip [denoted as D(1,1), D(2,2), D(3,3), D(4,4) in Fig. 12(a)], respectively. Similarly, in Figs. 14(a) and 14(b) we show the predicted spectra of the equivalent resonant orders for the 3D double-stub resonator. Similar to 3D single-stub resonators, the peak amplitude of 3D DS resonator is almost half of the 2D DS resonator and its contrast is lower compared to the 2D DS resonator.

**Table 3.2D Double-Stub Resonators**

| $L_1=L_2=L_{dip}$(nm) | FWHM (nm) | Q-factor | Finesse |
| --- | --- | --- | --- |
| 230 | 1182 | 1.38 | ----- |
| 710 | 515 | 3.16 | 2.04 |
| 1200 | 308 | 5.18 | 2.03 |
| 1670 | 217 | 7.23 | 2.07 |
| $L_1=L_2=L_{peak}$ (nm) | FWHM (nm) | Q-factor | Finesse |
| 450 | 821 | 1.92 | ----- |
| 930 | 397 | 3.9 | 1.90 |
| 1410 | 264 | 5.85 | 1.90 |
| 1890 | 199 | 7.73 | 1.87 |

*Spectrum specifications of first four resonant orders resulting in a dip or a peak at 1550nm in 2D DS resonatorpowertransmission.

**Table 4.3D Double-Stub Resonators**

| $L_1=L_2=L_{dip}$(nm) | FWHM (nm) | Q-factor | Finesse |
| --- | --- | --- | --- |
| 190 | ----- | ----- | ----- |
| 660 | 586 | 2.65 | 1.60 |
| 1140 | 339 | 4.56 | 1.68 |
| 1620 | 238 | 6.50 | 1.77 |
| $L_1=L_2=L_{peak}$ (nm) | FWHM (nm) | Q-factor | Finesse |
| 430 | 880 | 1.94 | 1.67 |
| 910 | 426 | 3.75 | 1.70 |
| 1390 | 288 | 5.46 | 1.72 |
| 1870 | 221 | 7.07 | 1.64 |

*Spectrum specifications of first four resonant orders resulting in a dip or a peak at 1550nm in 3D DS resonator power transmission.

Table 3 and Table 4 list the equal resonant length pairs that result in either a dip or a peak in the 2D and 3D DS resonators along with the specifications of their corresponding spectra obtained by scattering matrix formalism. The comparison reveals that similar to the single-stub resonators the resonant orders corresponding topeaks (at 1550nm) happen almost at the same stub lengths regardless of being 2D or 3D structures; however the resonant orders corresponding to dips (at 1550nm) happen at slightly different length pairs. Comparison of the results provided in Table 2for 3D SS resonator and Table 4 for 3D DS resonator reveals that single-stub and double-stub resonators show a peak in their transmission at the same stub lengths, however, double-stub resonators featuring a peak (dip) in their spectra have narrower (wider) FWHM and thus higher (lower)Q-factor compared to their corresponding single-stub resonators. The comparison of Fig.9(b) and Fig.14(b) also shows that double-stub resonators provide us with better contrast compared to single-stub resonators.

Starting from second resonant order of equal length pairs P(2,2) denoted as point A, in Figs. 15(a) and 15(b) and moving away from this peak point along the symmetry axis, which ensures the equality of the stub lengths, in Figs.15(c) and 15(d) we show how to tune the resonant wavelength in both 2D and 3D DS resonators, respectively. As expected, by increasing the stub lengths the resonant wavelength undergoes a red shift.

Our analysis shows that in both types of resonant orders (dips or peaks) in both 2D and 3D DS resonators, by increasing the stub length the FWHM of the resulting spectrum decreases. This again originates from the Fabry-Perot effect already explained in section A. It is worth noting that the same behavior occurs for a double-stub resonator of unequal stub lengths such that by keeping one stub length fixed andincreasing the length of the second stub the FWHM decreases (notshown). For instance, P(1,3) has narrower FWHM compared totheP(1,2). However, in a DS resonator of unequal stub lengths some extralobes and shoulders appear in the spectrum unlike in a DS resonator ofequal lengths. Therefore we inspected only DS resonators of equal stub lengths which produce neat and uniform spectra.

Further we found that, while tuning within each resonant order the spectrum FWHM remains almost unchanged, however the change in FWHM is far more from one resonant order to another.

Utilizing the transmission maps ofFigs.12(a) and 12(b) we can further locate the points that result in the plasmonic analogue to the electromagnetically induced transparency (PIT) phenomenon[19].To observe this phenomenon we should start from an equal stub length pair with the DS resonator and the corresponding SS resonators simultaneously having a dip in the transmitted power[36]. These points are depicted with white dots in Figs. 16 (a) and 16(b) for 2D and 3D DS resonators, respectively. The PIT effect is observed by breaking the symmetry of the DS resonator which is possible by making stub lengths slightly unequal while keeping the total length constant. Asymmetric DS resonators of slightly unequal stub lengths allow the formation of an asymmetric mode in the combined resonator of length $L_1+w+L_2$ ($w$ being the width of the waveguide) in addition to the symmetric mode and paves the way for their coupling which is the origin of the Fano-shaped transparency window in the PIT phenomenon[18].

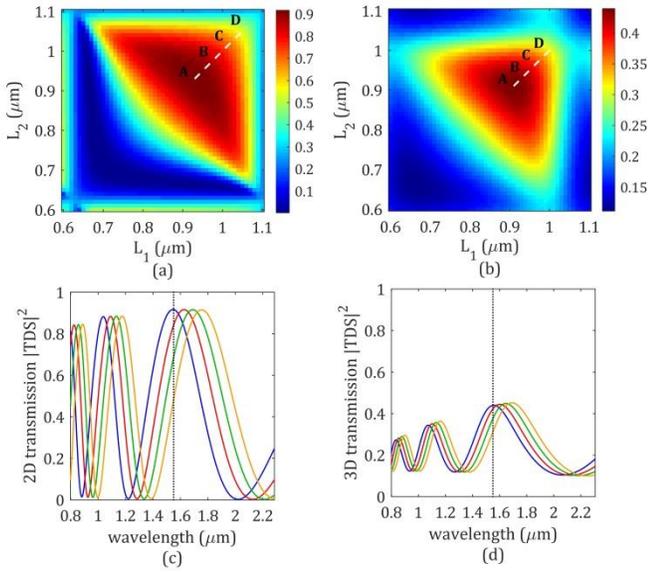

Fig.15. Four sets of equal-length pairs chosen from second resonant order along the symmetry axis of the transmission map of (a) 2D DS resonator, (b) 3D DS resonator. Power transmission spectra of (c) 2D DS resonator for length pairs A: $L_1=L_2$=930nm(blue line, $\lambda_{peak}$= 1550nm), B: $L_1=L_2$=980nm(red line, $\lambda_{peak}$=1626nm), C: $L_1=L_2$=1020nm(green line, $\lambda_{peak}$=1690nm), D: $L_1=L_2$=1060nm(yellow line, $\lambda_{peak}$=1752 nm), (d) 3D DS resonator for length pairs A: $L_1=L_2$=910nm (blue line, $\lambda_{peak}$=1550nm), B: $L_1=L_2$=940nm(red line, $\lambda_{peak}$=1600nm), C: $L_1=L_2$=970nm(green line, $\lambda_{peak}$=1650nm), D: $L_1=L_2$=1000nm(yellow line, $\lambda_{peak}$=1690nm).

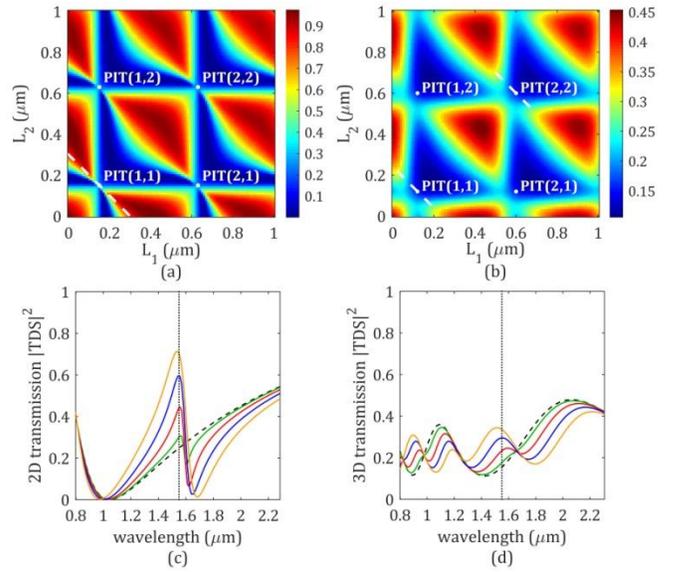

Fig.16. Asymmetric PIT-type spectra (a) 2D DS resonator (black dashed: $L_1=L_2$=(150,150)nm, green: (160,140)nm, red: (170,130)nm, blue: (180,120)nm, yellow: (190,110)nm, (b) 3D DS resonator(black dashed: $L_1=L_2$=(600,600)nm, green: (630,570)nm, red: (660,540)nm, blue:(690,510)nm, yellow: (720,480)nm.

Taking into account the fact that the resonant wavelength in the PIT phenomenon, for which a peak in the transmission appears, is proportional to the combined resonator length, by increasing one stub length by an amount of $dL$ and decreasing the other stub length by the same amount we can keep the total resonator length and thus PIT resonant wavelength fixed. This is possible by moving along the dashed white lines [Figs.16(a) and 16(b)] which pass through PIT(1,1) and PIT(2,2) resonant orders and make an angle of 45° with horizontal and vertical axes.

Starting from PIT(1,1) and moving along dashed white line [Fig. 16(a)] with steps of $dL$=10nm, in Fig. 16(c) we show the spectra of the 2D DS resonator for four sets of unequal length pairs. The dashed black curve gives the power transmission of the 2D DS resonator of equal stub lengths which due to the finite and rather wide width of our reference plasmonic MDM waveguide is not zero at 1550nm [18, 36].We found that by increasing $dL$ the asymmetry factor remains the same but the transmission amplitude increases[18, 36].

Starting from PIT(1,1) for 3D DS resonators and moving along dashed white line[Fig. 16(b)] we tried the same approach in 3D DS resonator, however we could not observe any reasonable asymmetric Fano-shaped transparency window for this PIT resonant order. We tried PIT(2,2) and the resulting spectra for four different sets of unequal stub lengths with steps of $dL$=30 nm are shown in Fig. 16(d). We conclude that observing PIT in 3D DS resonators is almost impossible which might be due to the open nature of 3D slot waveguide that does not allow the formation of well-defined junction resonator modes [18].

Although we could not observe PIT phenomenon in the studied 3D double-stub resonator, the PIT phenomenon has been observed in slightly decoupled stub pairs attached to a U-shaped 3D plasmonic waveguide[37].

A comparison of the reflection and transmission coefficients of the 2D geometries of Figs. 4(a)-4(d) with their 3D counterparts (results not shown) reveal that the $r_2$ coefficient—which is the reflection from a terminated waveguide end—is far less in the 3D terminated-end than in the 2D one, compared to all the other reflection and transmission coefficients. This is a sign of significant radiation loss in the 3D terminated ends. One way for increasing the $r_2$ coefficient is to increase the height of the metallic layer which converts the 3D slot-waveguide to MDM waveguide or adding block reflectors as high as 1μm to the terminated ends [38, 39]. Nonetheless, the latter approach not only adds complexity to the fabrication of the terminated ends but it also increases the device footprint.

At the terminated ends, the 3D slot waveguide is abruptly converted to a dielectric-metal-dielectric DMD waveguide. Due to the mismatch between the modal shapes of the slot-waveguide and DMD waveguide the SPP wave partially scatters to the cladding and substrate layers and partially is coupled to the MD and DMD modes. Therefore, one way to boost the $r_2$ reflection coefficient is to add a curved grating to the terminated end to reflect back the portion of the mode which is coupled to the DMD waveguide modes. We adapted the curved grating design in [40] and we saw some improvement in the $r_2$ coefficient and a 10% increase in the transmitted power of the double-stub resonator (results not shown). Regardless of how well the curved grating in a DMD waveguide is designed, due to the finite height of the grating blades the reflected SPP wave will suffer from radiation loss as it propagates through them. Moreover, the addition of a grating increases the device footprint which is a detriment.

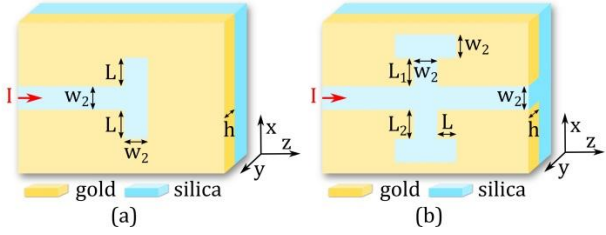

Fig.17. (a)Terminated end with a DS resonator,(b) double-stub resonatorwith DS resonator at the end of its stubs.

Another method that we tried for boosting the $r_2$ reflection coefficient was to add a double-stub resonator to the terminated end as shown in Fig. 17(a). According to the transmission map of the 3D DS resonator [Fig.12(b)] the length pairs that result in a dip in thetransmission of the DSresonator result in a peak in the reflection spectrum. Hence, by adding a double-stub resonator with a dip in its transmission and thus a peak in its reflection to the terminated end we can expect an increase in the $r_2$ reflection coefficient.

In Fig. 18(a) we compare the $r_2$ reflection coefficient, evaluated with numerical FDTD simulations, for 2D and 3D terminated ends along with the $r_2$ reflection coefficient of a terminated-end with a DS resonator [Fig.17(a)] with stub lengths $L$=190 nmcorresponding to the first order dip $D(1,1)$ shown in Fig. 12(b). As is seen, the $r_2$ reflection coefficient for 2D terminated-end is a lot higher than the 3D terminated end and the 2D terminated-end acts almost like a perfect mirror. Furthermore, it is evident that adding a DS resonator to the 3D terminated end increases the $r_2$ reflection coefficient. While$r_2$ is designed to be maximum at 1550 nm, due to the finite width of the reflection maximum of a DS resonator with $L$=190 nm stubs, there is a dip at 980nm in the $r_2$ coefficient of the DS terminated-end [green curve in Fig.18(a)].

Schematic of a double-stub resonator with a double-stub at the end of its stubs are shown in Fig. 17(b). Any change in the $r_2$ reflectioncoefficient affects the transmission map of the double-stubresonatorsuch that the locations of the resonant length pairs that correspond to dips or peaks change. By reevaluating the transmission map for a double-stub resonator (with double-stub terminations at the end of its stubs) we selected the length pairs$L_1$=$L_2$=680nm corresponding to the second-order peak $P(2,2)$ in its power transmission map (not shown) to be able to compare its transmitted power spectrum with the transmitted power spectrum of a double-stub resonator (without double-stub terminations at the end of its stubs) with stub lengths $L_1$=$L_2$=910nm corresponding to second-order peak in its power transmission map. The comparison of the two cases [Fig. 18(b)] reveals that the addition of DS resonator to the stub ends improves the spectrum contrast as well as the transmission amplitude.

## 6. CONCLUSIONS

In this work we compared resonators made out of 3D slot waveguides with their 2D MDM waveguide counterparts. To that end, we analyzed plasmonic single-stub and double-stub resonator geometries.

We have shown that scattering matrix theory can be applied to open 3D slot-waveguide based devices and we verified the predicted features with numerical FDTD simulations. This finding provides us with a powerful tool for predicting and manipulating the spectra of the 3D resonators by selecting the right stub lengths without the need to sweep over stub dimensions via full 3D FDTD simulations.

We found that despite the existence of radiation loss which degrades the performance of 3D structures, by a careful selection of resonator dimensions, 3D structures can have characteristics similar to the ones of 2D structures. We investigated designs for reducing radiation loss in 3D slot waveguides. We provided a double-stub based waveguide termination design which enhanced the resonator properties.

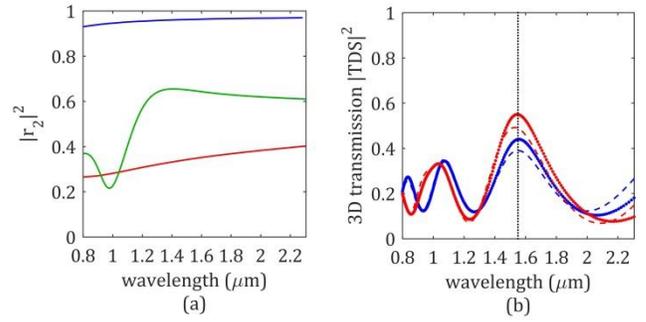

Fig.18. (a) Comparison of the reflection coefficients of terminated ends (TE). Blue line: 2D TE, red line: 3D TE, green line: 3D TE with DS at its end; (b) Comparison of the $|TDS|^2$ results. Blue curve: P(2,2) resonator as in Fig. 3(d), red curve: P(2,2) resonator with DS terminated stubs as in Fig 17(b). Solid lines are scattering matrix results and dashed lines are calculated numerically with 3D FDTD simulations.

The relatively thin metallic layer(115 nm) and the uniform silica ambient (n=1.44) of the 3D resonators lead to rather low Q values.However, filling the stubs with a high index material can boost the quality factor and improve the contrast of the 3D resonatorsby reducing the radiation loss [34]. Furthermore, increasing the thickness of the metal to 1000 nm or more reduces the radiation losssignificantly [27].Stub-resonators with low Q but with high concentration of fields can be of use in detector designs in optical interconnect applications that require very low energy consumption [41].

Plasmonic sensors usually have low Q values, however,they are very sensitive to the refractive index profile surrounding them and operate by detecting the change in the resonance frequency due to theperturbation of the local refractive index [42]. A bio-sensor based on a 3D plasmonic slot-waveguide has already been proposed [43], the addition of stub resonators integrated in slot waveguides can further boost the sensor's figure of merit by providing extra knobs to shape the spectral properties. The 3D double-stub resonators can further be utilized in the investigation of light-matter interactions in quantum plasmonic applications via positioning colloidal quantum dot emitters in hotspots within the DS resonator [39].

**Funding Information.** Scientific and Technological Research Council of Turkey (TUBITAK) Grant No: 112E247